\documentclass[journal]{IEEEtran}

\usepackage{amsfonts}
\usepackage{amssymb}
\usepackage{stfloats}
\usepackage{cite}
\usepackage{graphicx}
\usepackage{epstopdf}
\usepackage{psfrag}
\usepackage{subfigure}
\usepackage{amsmath}
\usepackage{array}
\usepackage{stfloats}
\usepackage{multirow}
\usepackage{color}
\usepackage{booktabs}
\usepackage{url}
\usepackage{graphicx}
\usepackage{enumitem}

\newtheorem{Thm}{Theorem}

\newtheorem{Def}{Definition}

\begin{document}

\title{An Analytical Framework for Delay Optimal Mobile Edge Deployment in Wireless Networks}

\author{Tao~Yu, Shunqing~Zhang,~\IEEEmembership{Senior Member, IEEE,} Xiaojing~Chen, and Shugong~Xu,~\IEEEmembership{Fellow, IEEE}
\thanks{This work was supported by the National Natural Science Foundation of China (NSFC) Grants under No. 61701293, No. 61871262, No. 61901251, and the National Key Research and Development Plan of China under No. 2017YFE0121400.}
\thanks{Tao Yu, Shunqing Zhang, Xiaojing Chen, and Shugong Xu are with Shanghai Institute for Advanced Communication and Data Science, Key laboratory of Specialty Fiber Optics and Optical Access Networks, Shanghai University, Shanghai, 200444, China (e-mail: \{yu\_tao, shunqing, jodiechen, shugong\}@shu.edu.cn).}
\thanks{Corresponding Author: Shunqing Zhang.}
}


\maketitle

\begin{abstract}
The emerging edge caching provides an effective way to reduce service delay for mobile users. However, due to high deployment cost of edge hosts, a practical problem is how to achieve minimum delay under a proper edge deployment strategy. In this letter, we provide an analytical framework for delay optimal mobile edge deployment in a partially connected wireless network, where the request files can be cached at the edge hosts and cooperatively transmitted through multiple base stations. In order to deal with the heterogeneous transmission requirements, we separate the entire transmission into backhaul and wireless phases, and propose average user normalized delivery time (AUNDT) as the performance metric. On top of that, we characterize the trade-off relations between the proposed AUNDT and other network deployment parameters. Using the proposed analytical framework, we are able to provide the optimal mobile edge deployment strategy in terms of AUNDT, which provides a useful guideline for future mobile edge deployment.
\end{abstract}


\IEEEpeerreviewmaketitle

\section{Introduction} \label{sect:intro}
\IEEEPARstart{A}{} recent paradigm shift of wireless traffics is the rapid growing interactive multimedia services, such as video clips, live streaming, and interactive games. Different from the conventional unidirectional video streaming, the interactive nature of aforementioned services usually require lower transmission delay on top of high data rate requirement. Since the centralized cloud based processing usually incurs significant delay and power consumption on the backhauling systems, the latest promising solutions to extend the caching capability towards the network edge side become quite popular \cite{mao2017survey}. By leveraging the storage and computing capabilities at the edge hosts (EHs), part of the information flow can be shared within local regions, and both the transmission delay and power consumption performance can be improved as well \cite{sengupta2016cache,gabry2016energy}.

A promising policy to enjoy the caching benefits in wireless communication areas is via a jointly design of the caching deployment strategy and the wireless information sharing scheme. For example, in a fully connected network scenario, various cache and delivery schemes across both backhaul and wireless transmission are proposed to minimize the transmission delay in terms of normalized delivery time (NDT) under centralized or decentralized manner \cite {tandon2016cloud,xu2018fundamental, girgis2017decentralized}. Similar ideas have also been extended to partially connected networks \cite{lampiris2019wyner,xu2019cache}. In \cite{lampiris2019wyner}, a larger degree of freedom is shown to be achievable through flexible backhaul configuration, and in \cite{xu2019cache}, a coded cooperation strategy can be utilized to achieve better NDT when base stations (BSs) and user terminals are equipped with caching capability.


The aforementioned works are based on the assumption that the signaling overhead of cooperative transmission and the EH deployment cost can be ignored. In the practical network planning, a more reasonable deployment strategy is to consider a shared edge caching policy, where each EH is able to connect to multiple BSs \cite{7931566}. With this strategy, the total number of EHs can thus be reduced and the cooperative transmission among multiple BSs is more convenient within each EH. However, there is limited work to consider the delay performance under this network topology, where the main challenges are listed as follows. First, the collaboration overhead has been ignored in the existing literature, and the information exchanges across different BSs are usually significant \cite {tandon2016cloud, xu2019cache}. Second, the conventional NDT performance focuses on the fully collaborated system by considering homogeneous transmission requirements and activities. In the practical scenario, the collaborative transmission is highly depending on the partial cooperation strategies, the network topology, as well as the user preference.  

To address aforementioned problems, we consider mobile edge enabled cooperative transmission system in this letter, where only intra-EH BS cooperation is permitted. In order to deal with the heterogeneous transmission requirements, we separate the entire transmission into backhaul and wireless phases, and propose {\em average user NDT (AUNDT)} as the performance metric. On top of that, we characterize the trade-off relations between the proposed AUNDT and other network deployment parameters. Using the proposed analytical framework, we are able to provide the optimal mobile edge deployment strategy in terms of AUNDT, which provides a useful guideline for future mobile edge deployment.

\section{System Model} \label{sect:sys_model} 

\begin{figure*}[t] 
\centering  
\includegraphics[height=6.5cm,width=14cm]{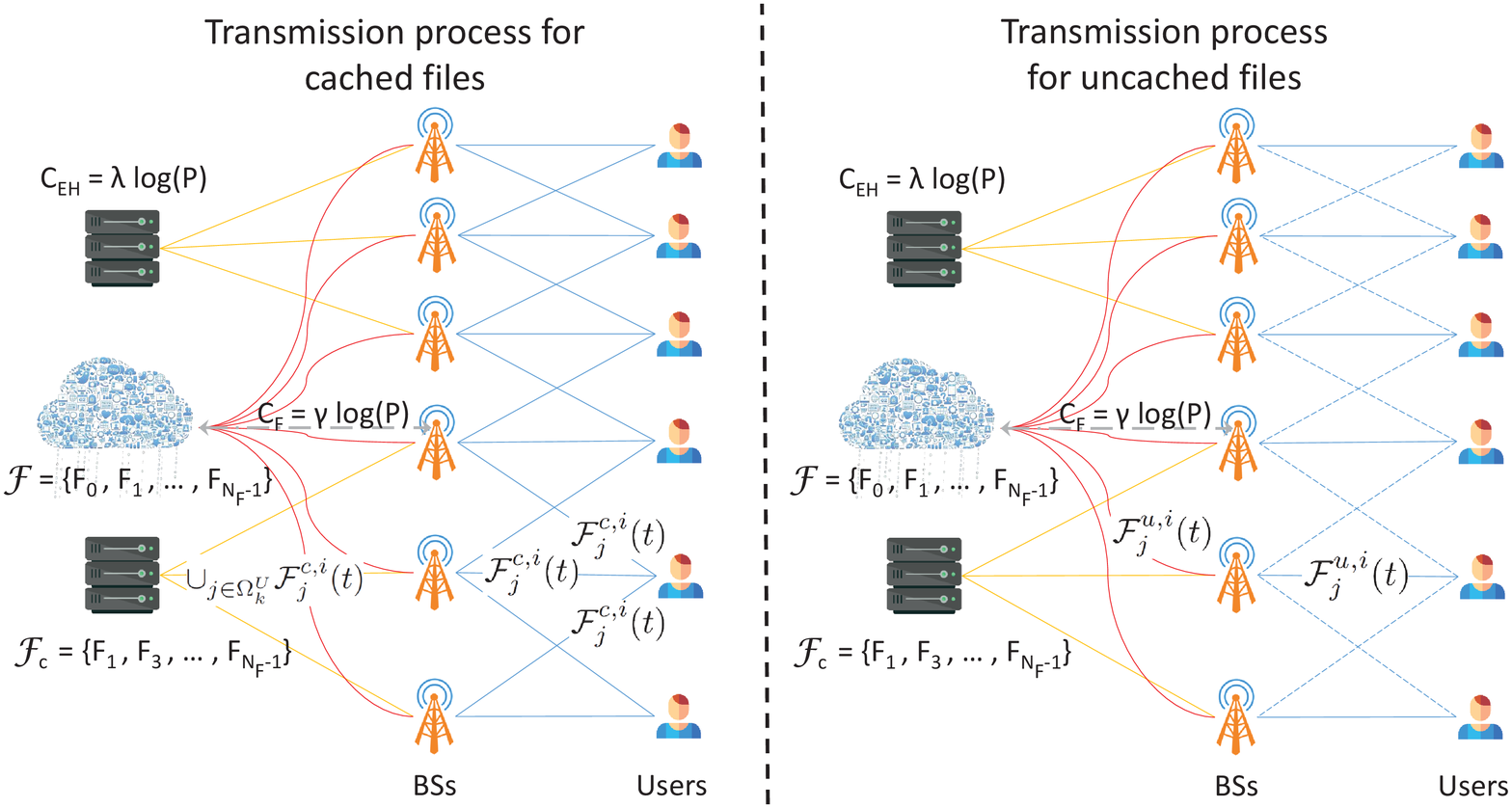}
\caption{A hypothetical mobile edge network, where cached files are transmitted to the BSs through the EH-BS links, and a BS cluster connected to the same EH can cooperatively transmit cached files to the users. Uncached files are first transmitted to the BSs through the Cloud-BS links, and then delivered to the users through interference channels by the BSs.}
\label{scenario}
\end{figure*}

Consider a cache-enabled partially connected wireless network with one cloud, $N_{EH}$ EHs, $N_{BS}$ BSs, and $N_{U}$ users as shown in Fig.~\ref{scenario}. Due to the sparse deployment of EHs, each EH is connected to one BS cluster with $M = N_{BS}/N_{EH}$ BSs\footnote{ $M$ is assumed to be one in the previous literature \cite{xu2018fundamental}, which means each BS has a unique serving EH. Although $M$ can be different from one EH to another, we leave it for our future works due to the page limit.}, since one EH can serve multiple BSs simultaneously. In addition, each user is able to retrieve interested contents via no more than $L$ ($\leq M$) consecutive BSs within the BS cluster. Denote $i \in \{1, \ldots, N_{EH}\}$ to be the index of EH or BS cluster, and $\Omega_i = \{M(i-1)+1, \ldots, Mi\}, \forall i$ to be the entire BS set connected to the $i^{th}$ EH, respectively. Let $\mathcal{F} = \{F_0, \dots, F_{N_F-1}\}$ represent the entire $N_F$ files available at the central cloud. The $i^{th}$ EH can only store a subset of files at the cloud, denoted as $\mathcal{F}_{EH}^{i} \subset \mathcal{F}$. We have $|\mathcal{F}_{EH}^{i}| = \mu N_{F}$, with $0 \leq \mu \leq 1$ denoting the ratio of the entire files stored in the $i^{th}$ EH.

For illustration purpose, we normalize the maximum transmit power of each BS, each EH, and the central cloud to be $P$, such that the maximum achievable rate of BS-user link, EH-BS link, and Cloud-BS link can be characterized by $C_{W} = \beta \log(P)$, $C_{EH} = \frac{\lambda}{M} \log(P)$, and $C_F = \gamma \log(P)$ bits/second per channel use, respectively \footnote{ To maintain the target error rate in the practical communication system, the working signal-to-interference and noise ratio (SINR) regions are similar on wireless \cite{shafik2006extended} and backhaul (optical) links \cite{freude2012quality}. Even if there are some SINR gaps, we incorporate this effect into the parameters $\gamma$ and $\lambda$ as well.}. With the above setting, the entire content transmission is divided into two phases, namely {\em the backhaul transmission phase} and {\em the wireless transmission phase}, as described below. 
\begin{itemize}
    \item{\em Backhaul Transmission Phase}: Each BS is assumed to maintain the cached content list of its serving EH. During the backhaul transmission phase, the content request of user $j$ at time slot $t$, $\mathcal{F}_j(t)$, is transmitted to the serving BSs. Upon receiving the content request, the serving BSs compare with their local content list, and divide it into two parts, i.e., the cached files, $\mathcal{F}^{c,i}_j(t) = \mathcal{F}_j(t) \cap \mathcal{F}_{EH}^{i}$, and the remaining files, $\mathcal{F}^{u,i}_j(t) = \mathcal{F}_j(t) - \mathcal{F}^{c,i}_j(t)$. The cached files can be retrieved from the $i^{th}$ EH directly, and the remaining files need to be obtained from the central cloud in a serial manner.
    \item{\em Wireless Transmission Phase}: The entire wireless transmission phase is divided into two stages, including the cooperative transmission stage and the interference transmission stage. During the cooperative transmission stage, all the serving BSs connected to the same EH, $\{\Omega_{i}\}$, transmit the information to the connected users via coded cooperative scheme as elaborated in \cite {xu2019cache}. During the interference transmission stage, each BS selects one user only and all the BSs deliver the information to their users via an interference channel environment.
\end{itemize}

The following assumptions are made throughout this letter. First, we consider $N_{U} = N_{BS}$ users are actively requesting their contents and BSs and users are equipped with only one antenna\footnote{When BSs and users are equipped with multiple antennas, we can apply the existing zero-forcing precoding \cite{yoo2005optimality} or the coordinated beamforming technique \cite{chae2012interference} to eliminate the inter-stream interference. Hence, we can treat the MIMO scenario as multiple BS-user pairs for simplicity.}. Second, we consider the linear deployment of BSs as shown in Fig.~\ref{scenario}, and all the BSs connected the same EH share the same frequency band in order to facilitate the cooperative transmission. In addition, the entire frequency band is divided into two orthogonal sub-bands and alternatively assigned to BS clusters, such that any two neighboring BS clusters are not interfering with each other. Third, the BSs connected the neighboring EHs utilize orthogonal frequency bands to avoid the potential interference, and the cooperations of BSs belonging to different EHs are not allowed. Last but not least, user preferences are uniformly distributed, i.e., all users request each file with the same probability.

\section{Transmission Delay and Problem Formulation} \label{sect:problem_formulation} 

In this section, we briefly calculate the overall transmission delay of the above system and define the average user normalized delivery time (AUNDT) as the proposed performance metric. 

\subsection{Overall Transmission Delay}
Due to the symmetric property of the linear network deployment, we focus on the BS cluster connected to the $i^{th}$ EH and the corresponding users as shown in Fig.~\ref{scenario} to describe the entire transmission process. 

\subsubsection{Backhaul Transmission Phase} During the backhaul transmission phase, the $k^{th}$ BS ($k \in \Omega_i$) collects the content requests of $\mathcal{F}_{j}(t)$ from all the connected serving user set, $\Omega^{U}_k$. In order to facilitate the cooperation in the wireless transmission phase, it requests the combined information $\cup_{j \in \Omega^{U}_k} \mathcal{F}^{c,i}_j(t)$ from the $i^{th}$ EH. Since only one user can be served during the interference transmission stage, we simply choose user $j = k$ and request the remaining file for user $j$, $\mathcal{F}^{u,i}_{j}(t)$. With the above notations, the backhaul transmission delay is given by,
\begin{eqnarray}
T_{k}^{BH}(t) = \frac{ \zeta \left(\cup_{j \in \Omega^{U}_k} \mathcal{F}^{c,i}_j(t)\right) }{ C_{EH} } + \frac{ \zeta \left ( \mathcal{F}^{u,i}_{k}(t) \right ) }{ C_F },
\end{eqnarray}
where $\zeta(\cdot)$ is a function to return the length of files.

\subsubsection{Wireless Transmission Phase} 

The entire wireless transmission phase contains the cooperative transmission stage and the interference transmission stage. In the cooperative transmission stage, each BS cluster serves its connected users via the coded cooperative scheme proposed in \cite {xu2019cache}. In the interference transmission stage, each BS only serves the dedicated user (e.g., with the same index) via the interference channel. In this sense, the delay of wireless transmission phase can be given by,
\begin{eqnarray}
T_{k}^{W}(t) = \frac{ \zeta \left(\cup_{j \in \Omega^{U}_k} \mathcal{F}^{c,i}_j(t)\right) }{ C_{WC,k} } + \frac{ \zeta \left ( \mathcal{F}^{u,i}_{k}(t) \right ) }{ C_{WB} },
\end{eqnarray}
where $C_{WC,k}$ and $C_{WB}$ are the achievable throughput in the cooperative transmission stage and the interference stage, respectively. With the above formulation, the overall transmission delay for the user $j=k$ is thus given by,
\begin{equation} 
\label{eq:total_delay}
T_k(t) = T_{k}^{BH}(t) + T_{k}^{W}(t).
\end{equation}

\subsection{Problem Formulation}

In order to measure the asymptotic delay performance of the EH enabled system, NDT is commonly used, either in the fully connected architecture \cite {tandon2016cloud} or in the partially connected network with homogeneous delay requirements \cite {xu2019cache}. Due to the partially connected nature of EHs, the achievable throughput in the cooperative transmission stage $C_{WC,k}$ is in general different. Here we introduce a more reasonable performance metric, \emph{AUNDT}, to characterize the heterogeneous delay behaviors of different users. 

\begin{Def}[AUNDT] Given the normalized EH cache capacity $\mu$, the normalized rates of BS-user link, EH-BS link and Cloud-BS link, i.e., $\beta, \frac{\lambda}{M}$ and $\gamma$, the NDT of user $j$ under asymptotically large file length $Z$ and transmit power $P$ is defined as, 
\begin{equation} \label{eq:ndt_user}
\begin{split}
\tau_j(\mu, \gamma, \lambda, \beta, M ) 
&\triangleq {\lim_{P \to \infty}} \lim_{Z \to \infty} \sup \frac{ T_j(t) }{Z/\log P}. 
\end{split}
\end{equation}
By averaging the achievable NDTs of all users, we can define \emph{AUNDT} as,
\begin{equation} \label{eq:aundt}
\begin{split}
\tau(\mu, \gamma, \lambda, \beta, M )
&\triangleq  \frac{\sum_{j = 1}^{N_U}\tau_j(\mu, \gamma, M, \lambda)}{N_U}.
\end{split}
\end{equation}
\end{Def}

AUNDT represents the averaged asymptotic delay performance of all possible users in terms of NDT\footnote{We can introduce the weighted NDT as well, if user priorities need to be taken into consideration.}. The optimal EH deployment strategy can be obtained by solving the following AUNDT minimization problem.
\begin{eqnarray}
\mathop{\textrm{minimize}}_{M} && \tau(\mu, \gamma, \lambda, \beta, M), \nonumber \\
\textrm{subject to} && M \geq L, M \in \mathbb{Z}^{+}. 
\label{eq:opti}
\end{eqnarray}

The above formulation is a typical mixed integer optimization problem, which in general difficult to solve. Furthermore, since the objective function $\tau(\mu, \gamma, \lambda, \beta, M)$ highly depends on the users' instantaneous preference, e.g., $\{\cup_{j \in \Omega^{U}_k} \mathcal{F}^{c,i}_j(t)\}$, a brute force evaluation of $\tau(\mu, \gamma, \lambda, \beta, M)$ is computationally prohibited.

\section{Proposed Deployment Strategy and Analysis} \label{sect:deployment_strategy} 

In order to obtain a mathematically feasible solution, we focus on the case where each user requests different files\footnote{As a matter of fact, when users request some similar files, the BS combines similar requests and the number of required files for EH-BS link can be reduced, which results in a better AUNDT. }, such that $|\cup_{j \in \Omega^{U}_k} \mathcal{F}^{c,i}_j(t)| = \sum_{j \in \Omega^{U}_k} |\mathcal{F}^{c,i}_j(t)|$ for all possible $t$ and $j$. With this assumption, we can derive the optimal deployment density $M^{\star}$ and analyze the deployment strategy in what follows. 

\subsection{Optimal Deployment Density}
\label{subsect:deployment_density} 

By applying the convex relaxation to the constraint $M \in \mathbb{Z}^{+}$, the original AUNDT minimization problem~\eqref{eq:opti} can be well approximated by a standard convex optimization problem. With some mathematical manipulations as shown in the proof, we can have the optimal deployment density given as stated in following theorem.

\begin{Thm}
\label{thm_opt_dep}
Given the normalized EH cache capacity $\mu$, the normalized rates of BS-user link, EH-BS link and Cloud-BS link, i.e., $\beta, \lambda$ and $\gamma$, the AUNDT performance is given by, 
\begin{eqnarray}
\begin{split}
\tau(\mu, \gamma, \lambda, \beta, M) &=  \mu \left ( \frac{ M L}{\lambda} + \frac{2G(L)}{\beta M} - \frac{1}{\gamma} - \frac{2}{\beta} \right) \\ 
&+ \frac{1}{\gamma} + \frac{4}{\beta},
\end{split}
\end{eqnarray}
and the optimal deployment density of EHs, $M^{\star}$, is given by,
\begin{eqnarray}
M^{\star} & = & 
\max \left \{ L, \left \lceil \sqrt{ \frac{2 \lambda}{\beta L} G(L) } - \frac{1}{2} \right \rceil \right \}, \text{for $\mu > 0$}, 
\end{eqnarray}
where $G(L)= 1-L + 4L\sum_{j = 1}^{\frac{L-1}{2}}\left ( \frac{1}{L+2j-1}\right)$ is the associated delay performance loss when the cooperation of cross-EH BSs is not allowed\footnote{When the cooperation of cross-EH BSs is allowed, $G(L) = 0$ and we can thus have the lower bound of AUNDT.}.
\end{Thm}

\IEEEproof
Please refer to Appendix~\ref{appendix:proof_thm1} for the proof.
\endIEEEproof

From Theorem~\ref{thm_opt_dep}, we can conclude that the optimal deployment density is highly related to the normalized EH-BS  rate $\lambda$ and BS-user rates $\beta$, and is independent of the normalized Cloud-BS rate $\gamma$ as well as the caching capacity $\mu$. This is because the optimal deployment strategy shall jointly consider the delivery capability provided by EHs and BSs, including the transmission capacity between EH and BSs, $\lambda$, and BS-user links, $\beta$. Meanwhile, since the derivative of $\tau(\mu, \gamma, \lambda, \beta, M)$ with respect to $M$ is independent of $\mu$ and $\gamma$, we can ignore them in the initial network planning stage.

\begin{figure}[t]
\centering  
\includegraphics[height=6.5cm,width=7cm]{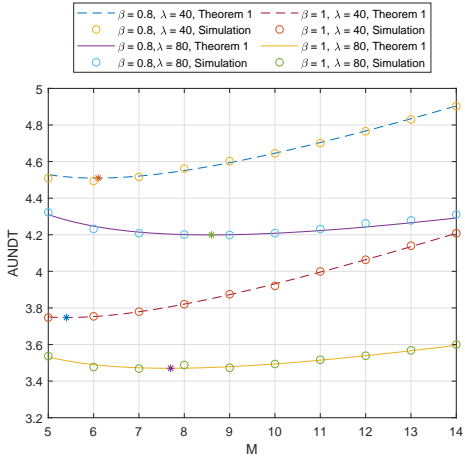}
\caption{AUNDT v.s. $M$ when $\mu = 0.7$ and $\gamma = 1.5$. A larger BS-user rate $\beta$ leads to a denser EH deployment and a lower AUNDT. A larger EH total rate $\lambda$ leads to a sparser deployment and a lower AUNDT. }
\label{fig_simu2}
\end{figure}

\begin{figure}[t]
\centering  
\includegraphics[height=6.5cm,width=7cm]{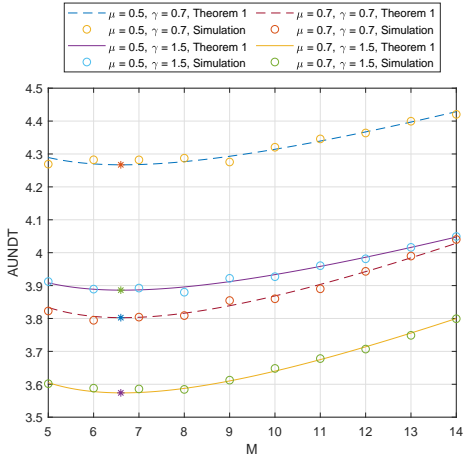}
\caption{AUNDT v.s. $M$ when $\beta = 1$ and $\lambda = 60$. Larger cache capacity $\mu$ and Cloud-BS rate $\gamma$ bring a lower AUNDT, but they do not affect optimal deployment strategy.}
\label{fig_simu1}
\end{figure}

In the following numerical simulation, we choose $L = 5$, $N_F = 500$, $N_{U} = N_{BS} = M$, $P = 20$w, and vary the values of $\mu, \gamma, \lambda, \beta, M$ to generate different curves as shown in Fig.~\ref{fig_simu2} and Fig.~\ref{fig_simu1}. The theoretical results from Theorem~\ref{thm_opt_dep} are plotted by dashed curves and compared with the numerical generated results (circles). As shown in these figures, the optimal deployment density shall jointly consider the values of $\lambda$ and $\beta$, while ignoring the values of $\mu$ and $\gamma$.

\subsection{Analysis of Deployment Strategy}
\label{subsect:analysis_unavailable}

In the practical cellular networks, the optimal deployment density $M^{\star}$ may not be achievable due to limited budget or space. In this situation, a more meaningful question is how to adjust EHs' parameters for better AUNDTs. By solving the inequalities of $\tau(\mu, \gamma, \lambda, \beta, M^{\star}) \geq \tau(\mu, \gamma, \lambda', \beta, M')$ and $\tau(\mu, \gamma, \lambda, \beta, M^{\star}) \geq \tau(\mu', \gamma, \lambda, \beta, M')$, we have the following theorem.

\begin{Thm}
\label{thm_para}
For any $M' \leq M^{\star}$ and fixed $\lambda, \beta, \gamma, \mu$, the AUNDT performance can still be guaranteed if the adjusted values of normalized EH-BS rate $\lambda'$ or EH cache capacity $\mu'$ shall satisfy the following relations.
\begin{eqnarray}
\lambda' & \geq & \frac{\lambda M'}{M^{\star} + \left( \frac{1}{M^{\star}} - \frac{1}{M'} \right) \frac{2 \lambda G(L)}{\beta L} }, \\
\mu' & \geq & \mu \left [ \frac{\frac{L}{\lambda}(M - M') + \left( \frac{2}{\beta M} - \frac{2}{\beta M'} \right) G(L)}{\frac{L M'}{\lambda} + \frac{2G(L)}{\beta M'} - \left(\frac{1}{\gamma} + \frac{2}{\beta} \right)} +1 \right]
\end{eqnarray}
\end{Thm}

\begin{figure}[t] 
\centering  
\includegraphics[height=6.5cm,width=7cm]{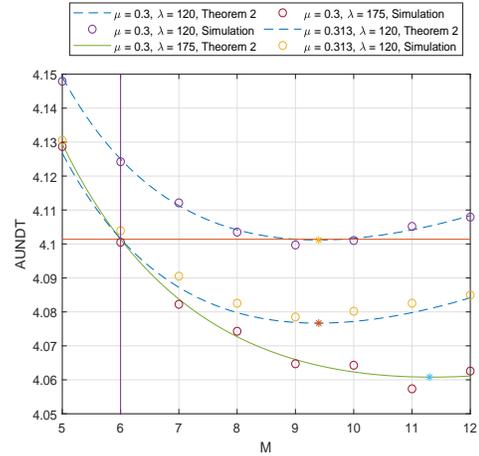}
\caption{AUNDT v.s. $M$ when $\beta = 1$ and $\gamma = 1.5$. When the optimal deploy density $M^* = 9$ is not available and the available deploy density $M' = 6$, adjusted cache capacity $\mu' = 0.313$ or adjusted EH-BS rate $\lambda = 175$ guarantee the same delay performance. }
\label{fig_simu3}
\end{figure}

In Fig.~\ref{fig_simu3}, dashed lines represent the AUNDT performance for given $\lambda, \beta, \gamma, \mu$. As shown in this figure, the optimal deployment density $M^{*}$ is equal to 9. If the achievable density $M'$ in the practical deployment is limited by 6, we can also adjust the cache capacity $\mu$ or normalized EH-BS rate $\lambda$ to guarantee the target AUNDT performance. 

\section{Conclusion} \label{sect:conc}

In this letter, we propose a new performance metric called AUNDT to deal with the heterogeneous transmission requirements by jointly considering the backhaul and wireless transmission phases. Moreover, a theoretical framework for partially connected edge cache network is derived to characterize the trade-off relations between AUNDT and deployment parameters, and the optimal deployment strategy is obtained thereafter. Through this work, we hope it can pave the way for future edge cache deployments. 

\begin{appendices} 
\section{Proof of Theorem~\ref{thm_opt_dep}} \label{appendix:proof_thm1}

\subsection{Derivation of AUNDT}

Assume that caching strategy of EHs is to cache each file with the equal probability. Note that $\mu$ characterizes the cache capacity of EH. By strong law of large numbers, $\mu$ can also represent the cached proportion of each file. Thus, we can get the length of $\mathcal{F}^{c,i}_j(t)$ and $\mathcal{F}^{u,i}_j(t)$,
\begin{equation}
\label{eq:file_length_1}
\zeta \left (\mathcal{F}^{c,i}_j(t) \right )= \mu Z, \quad \zeta \left (\mathcal{F}^{u,i}_j(t) \right )= (1-\mu) Z.
\end{equation}
Under the worst case where each user requests a different file, we have,
\begin{equation} 
\label{eq:file_length_2}
\zeta \left(\cup_{j \in \Omega^{U}_k} \mathcal{F}^{c,i}_j(t)\right) =  \sum_{j \in \Omega^{U}_k} \zeta \left (\mathcal{F}^{c,i}_j(t) \right )=\mu L Z.
\end{equation}
Define a function $\psi (k)$ to return the number of cooperative BS-user links of BS $k$. From the network topology, $\psi (k)$ is given by,
\begin{enumerate}[label=\roman*)]
\item $\psi (k) = k - M(i-1) + \frac{L-1}{2}$, if $M(i-1) < k \leq M(i-1) + \frac{L-1}{2}$, 
\item $\psi (k) = L$, if $ M(i-1) + \frac{L-1}{2} < k \leq Mi - \frac{L-1}{2}$, 
and \item $\psi (k) =Mi - k + \frac{L+1}{2}$, if $Mi - \frac{L-1}{2} < k \leq Mi$.
\end{enumerate}

Through the coding and transmission scheme for large transmitter cache size as elaborated in \cite {xu2019cache}, a degree of freedom (DoF) of $1$ can be achieved while all $L$ links are activated. In our scheme, since each BS-user link occupies the same transmission resources of a BS, we suppose that the achievable throughput entirely depends on the number of active BS-user link. Under the alternative orthogonal spectrum assignment scheme as illustrated in Section.~\ref{sect:sys_model}, the $C_{WC,k}$ can then given by,
\begin{eqnarray} \label{eq:coop_rate}
C_{WC,k} =
\frac{1}{2} \frac{ \psi (k) }{L} \beta \log P = \frac{ \beta\psi (k) }{2L} \log P.
\end{eqnarray}

For non-cooperative transmission, it is shown in \cite{el2014interference} that 
for partially connected network, the asymptotic per user DoF under an interference channel is $1/2$, and we can obtain,
\begin{eqnarray}
\label{eq:interference_rate}
C_{WB} = \frac{1}{4} \beta \log P.
\end{eqnarray}

Substituting \eqref{eq:file_length_1} - \eqref{eq:interference_rate} into \eqref{eq:ndt_user}, we have
\begin{eqnarray*}
\tau_j(\mu, \gamma, \lambda, \beta, M ) 
&= (1 - \mu) \left (\frac{1}{\gamma} + \frac{4}{\beta} \right) + \frac{\mu M L}{\lambda} +\frac{2\mu L}{\beta \psi (j)}.
\end{eqnarray*}

AUNDT can be obtained by only focusing on the BS cluster connected to the $i^{th}$ EH and the corresponding users due to the symmetric property of the linear network deployment,
\begin{eqnarray}
& & \tau(\mu, \gamma, \lambda, \beta, M ) =\frac{\sum_{j = 1}^{M}\tau_j(\mu, \gamma, M, \lambda)}{M} \nonumber \\
& = &\mu \left ( \frac{ M L}{\lambda} + \frac{2G(L)}{\beta M} - \frac{1}{\gamma} - \frac{2}{\beta} \right) + \frac{1}{\gamma} + \frac{4}{\beta}, \label{appendix:ndt}
\end{eqnarray}
where $G(L)$ is a function of $L$, defined as $G(L)= 1-L + 4L\sum_{j = 1}^{\frac{L-1}{2}}\left ( \frac{1}{L+2j-1}\right)$.

\subsection{Derivation of $M^*$}

The optimization of the deployment density $M$ in the expression of AUNDT (\ref{appendix:ndt}) is a user-behavior-independent mixed integer programming. To solve this problem, We apply convex relaxation to relax $M \in \mathbb{Z}^{+}$ into real domain $\mathbb{R}^{+}$. The optimization problem (\ref{eq:opti}) can hence be expressed as:
\begin{eqnarray}
\label{relaxed_problem}
\mathop{\textrm{minimize}}_{M} &&\mu \left ( \frac{ M L}{\lambda} + \frac{2G(L)}{\beta M} - \frac{1}{\gamma} - \frac{2}{\beta} \right) + \frac{1}{\gamma} + \frac{4}{\beta}, \nonumber \\
\textrm{subject to} && M \geq L, M \in \mathbb{R}^{+}.
\end{eqnarray}

Problem~(\ref{relaxed_problem}) is a convex problem. The optimal deployment $\widetilde{M}^*$ can be achieved when the derivative of the objective function $\tau(\mu, \gamma, \lambda, \beta, M )$ equals to $0$. If the optimal deployment $\widetilde{M}^* < L$, let $\widetilde{M}^* = L$. The closed-form solution to (\ref{relaxed_problem}) is then given by: 

\begin{equation}
\widetilde{M}^* = 
\begin{cases}
\max \left \{ L, \sqrt{ \frac{2 \lambda}{\beta L} G(L) } \right \}, & \text{if $\mu > 0$}, \\
\text{any value greater than $L$}, & \text{if $\mu = 0$}.\\
\end{cases}
\end{equation}

The solution to the original non-convex problem (\ref{eq:opti})  can be obtained by rounding $\widetilde{M}^*$, i.e. ,  $M^* = \left \lceil \widetilde{M}^* - \frac{1}{2} \right \rceil$.

\end{appendices}


\bibliographystyle{IEEEtran}
\bibliography{IEEEfull,references}

\end{document}